\documentclass[modern]{aastex631}

\newcommand\psp{\emph{PSP}}

\begin{document}

\title{Highly Polarized Type III Storm Observed with Parker Solar Probe}

\correspondingauthor{Marc Pulupa}
\email{pulupa@berkeley.edu}

\author[0000-0002-1573-7457]{Marc Pulupa}
\affiliation{Space Sciences Laboratory, University of California, Berkeley, CA 94720-7450, USA}

\author[0000-0002-1989-3596]{Stuart D. Bale}
\affiliation{Space Sciences Laboratory, University of California, Berkeley, CA 94720-7450, USA}
\affiliation{Physics Department, University of California, Berkeley, CA 94720-7300, USA}

\author[0000-0002-0606-7172]{Immanuel Christopher Jebaraj}
\affiliation{Space Research Laboratory, University of Turku, Turku, Finland}

\author[0000-0002-4559-2199]{Orlando Romeo}
\affiliation{Space Sciences Laboratory, University of California, Berkeley, CA 94720-7450, USA}

\author[0000-0002-2002-9180]{Säm Krucker}
\affiliation{University of Applied Sciences and Arts Northwestern Switzerland, Bahnhofstrasse 6, 5210 Windisch, Switzerland}
\affiliation{Space Sciences Laboratory, University of California, Berkeley, CA 94720-7450, USA}








\begin{abstract}

  The \emph{Parker Solar Probe} (\psp{}) spacecraft observed a large coronal mass ejection (CME) on 5 September 2022, shortly before closest approach during the 13th \psp{} solar encounter.  For several days following the CME, \psp{} detected a storm of Type III radio bursts. Stokes parameter analysis of the radio emission indicates that the Type III storm was highly circularly polarized. Left hand circularly polarized (LHC) emission dominated at the start of the storm, transitioning to right hand circularly polarized (RHC) emission at the crossing of the heliospheric current sheet on 6 September.

  We analyze the properties of this Type III storm. The drift rate of the Type IIIs indicates a constant beam speed of $\sim$0.1$c$, typical for Type III-producing electron beams. The sense of polarization is consistent with fundamental emission generated primarily in the $O$-mode.
  The stable and well organized post-CME magnetic field neatly separates the LHC- and RHC-dominated intervals of the storm, with minimal overlap between the senses of polarization. The proximity of \psp{} to the source region, both in radial distance and in heliographic longitude, makes this event an ideal case study to connect \emph{in situ} plasma measurements with remote observations of radio emission.


\end{abstract}




\section{Introduction} \label{sec:intro}

Type III solar radio emission is generated by beams of electrons accelerated near the Sun \citep{2014RAA....14..773R}. Electron beams accelerated on open magnetic field lines can generate interplanetary (IP) Type III emission at low frequencies, with descending frequency profiles that can extend to the local plasma frequency \citep{2000GMS...119..115D}.

Type III emission can occur as discrete individual bursts, or as a Type III storm, a quasi-continuous series of small bursts. Type III storms typically last for several days, and in some cases longer than a full solar rotation \citep{1970SoPh...15..222F,1970SoPh...15..433F,1971SoPh...17..392F}. \citet{2010ApJ...708L..95E} analyzed intensity and waiting time distributions for a Type III storm observed by STEREO and found statistics consistent with a Poisson process, featuring individual bursts occuring independently as a result of a persistent driver.

IP Type III storms are typically a low-frequency extension of ``metric'' Type III storms which occur at higher frequencies, with a solar active region as the common source for both the IP and metric storms. The persistence of storm emission is indicative of a stable active region configuration which is favorable for accelerating electrons into IP space \citep{1984A&A...136..255B,1984A&A...141...17B}. Reconfiguration of the magnetic field, for example by a coronal mass ejection (CME), can disrupt a Type III storm, causing cessation of the storm activity \citep{2001SoPh..204..121R,2004P&SS...52.1399G}.

\citet{2007ApJ...657..567M}, using Geotail and Akebono observations, suggested that Type III storms are generated near the boundary between an active region and the open field lines of a coronal hole.  \citet{2011A&A...526A.137D} analyzed Hinode/EIS and Nançay Radioheliograph data, and proposed that interchange reconnection at this type of boundary can produce coronal outflows and metric Type III storms.

Recently, \citet{9814233,9814336} has demonstrated a relationship between Type III storms and energetic particle flux associated with CIRs and CMEs, suggesting that the interchange reconnection that produces storms can also provide seed particles which can be accelerated to high energies by these propagating structures.

\section{Data}

The \emph{Parker Solar Probe} (\psp{}) spacecraft \citep{2016SSRv..204....7F} launched in 2018 into an elliptical orbit around the Sun, with a perihelion significantly closer to the solar surface than any previous spacecraft. \psp{} uses Venus gravity assist (VGA) maneuvers to successively lower its perihelion. After the seventh and final planned VGA on 6 November 2024, the spacecraft will reach a final perihelion distance of 9.86 solar radii ($R_\odot$) on 24 December 2024. \psp{} completed its thirteenth solar encounter (E13) during September 2022, with a perihelion distance of 13.3 $R_\odot$.

We use magnetic field and radio data from the \psp{}/FIELDS instrument suite \citep{2016SSRv..204...49B}. The magnetic field data are provided by the outboard fluxgate magnetometer (MAG) sensor mounted behind the spacecraft on the FIELDS magnetometer boom. Radio measurements are made with the FIELDS Radio Frequency Spectrometer (RFS) \citep{2017JGRA..122.2836P}. During E13 (as in previous encounters), the RFS measured two channels of auto- and cross-spectral data, using the $V1-V4$ FIELDS antennas, which are mounted near the edge of the \psp{} heat shield and protrude into the sunlight. One channel made measurements from the $V1-V2$ dipole, and the other channel from the $V3-V4$ dipole. A diagram of this `cross dipole' configuration can be found in \citet{2022A&A...668A.127P}.

The RFS produces data in two frequency ranges. The Low-Frequency Receiver (LFR) range covers a bandwidth of 10 kHz-1.7 MHz, and High-Frequency Receiver (HFR) data covers 1.3 MHz-19.2 MHz. During E13, both LFR and HFR spectra were produced at a cadence of approximately 3.5 seconds per spectrum.

The RFS receiver has two gain stages. During nominal operations, the receiver is set to automatically select between high and low gain. The RFS automatic gain selection algorithm \citep{2017JGRA..122.2836P} accumulates high and low gain spectra to produce LFR and HFR telemetered spectra, which typically consist of 40 averaged individual onboard FFTs. The 3.5 second cadence is achieved when the receiver can accumulate the target of 40 spectra in high gain. During sustained periods of continuous high amplitude plasma waves, the processing and accumulation of the low gain spectra in addition to the high gain spectra can slow the cadence of the receiver from approximately 3.5 seconds to 7 seconds per spectrum. The RFS returned low gain spectra at this lower cadence for approximately 16\% of the interval shown in Figure \ref{fig:spectrogram}.

All RFS data products in this paper use Level 3 RFS data. In Level 3 data, corrections are applied to account for higher noise background noise during low gain intervals, instrument noise from the preamplifier and analog components in the receiver is removed, and the antenna effective length determined by \citet{2022A&A...668A.127P} is used to convert spectral quantities to physical units of $\mathrm{W/m^2/Hz}$.

\subsection{Circular Polarization}

Using the auto and cross spectral data together allows us to construct Stokes parameters, which describe the intensity and polarization of the observed radio emission. We focus on the Stokes $V$ parameter, representing the circular polarization. Stokes $V$ is often presented normalized to the Stokes intensity $I$, where fully circularly polarized emission would have a $V/I$ value of -1 or 1. In the Level 3 RFS HFR and LFR data files, this normalized quantity is contained in the \texttt{STOKES\_V} variables.

Since launch, \psp{} has observed circularly polarized emission in individual Type III bursts \citep{2020ApJS..246...49P, 2023A&A...674A.105D,2023ApJ...955L..20J}, Type III storms \citep{2020ApJS..246...49P}, and Type II radio bursts driven by CMEs \citep{2022ApJ...930...88N}. Reported observations of polarized IP solar radio emission from other spacecraft, such as Wind \citep{1995SSRv...71..231B}, STEREO \citep{2008SSRv..136..487B}, and Cassini \citep{2004SSRv..114..395G}, are rare. This is likely due to a combination of factors: firstly, not all radio bursts show significant polarization \citep{2023A&A...674A.105D}. Additionally, the polarized component of emission seen on \psp{} is most significant above 1 MHz, and STEREO and Wind are often operated in a mode which only enables polarization measurements for frequencies less than $\sim$1 MHz.

We note that Stokes $V$ values used in this work result from an updated version of the simplified calculation described in \citet{2020ApJS..246...49P}. The value of $V$ is corrected for the effects of antenna non-orthogonality and non-equal antenna effective length using Mueller matrices, as described in \citet{2011pre7.conf...13L}. We use the convention where Stokes $V>0$ corresponds to right hand circularly polarized emission (RHC), while $V<0$ corresponds to left hand circular emission (LHC). We note that the works mentioned in the previous paragraph used the opposite convention. This change was implemented during the development of the Level 3 RFS data, in order to be consistent with the IAU definition of the Stokes parameters \citep{1996A&AS..117..161H,2021hai1.book..127R}. We emphasize that the physical interpretation of the wave handedness has not changed, merely the sign convention for Stokes $V$.

\section{Observations}

\subsection{Context: Encounter 13 CME}

Shortly before the E13 perihelion in early September 2022, \psp{} observed a complex interplanetary coronal mass ejection (ICME). Remote sensing and in situ observations of the CME are described in \cite{2023ApJ...954..168R}, hereafter R23. We present here a highly compressed description of the CME timeline described in detail in R23, to provide the broader context for our subsequent focused discussion of the Type III storm. All times discussed below are UTC.

Remote sensing observations of the CME eruption commence at 09-05/16:10 ($t_0$ in R23), and the ICME shock arrives \emph{in situ} at 09-05/17:27 ($t_1$ in R23). After the shock passage, \psp{} enters the ICME sheath, followed by a period of closed field lines. This period features rapidly changing magnetic fields and bi-directional electron streaming. Following this dynamic period, the field becomes more stable, with predominantly outward field from 09-06/03:26 to 09-06/17:28 ($t_{5b}$ and $t_{7a}$ in R23). This period of outward field includes a period of typical near-Sun solar wind, and a lower density period of sub-Alfvénic flow ($M_A < 1$).

After the period of outward field, the spacecraft then crosses a current sheet into a region of steady inward field, starting at 09-06/17:40 ($t_{7b}$ in R23).
Plasma conditions change abruptly at the current sheet crossing, with density dropping from ${>}1000\:\text{cm}^{-3}$ to ${\sim}100\:\text{cm}^{-3}$ and velocity decreasing from ${\sim}350\:\text{km\:s}^{-1}$ to ${\sim}200\:\text{km\:s}^{-1}$, leading to highly sub-Alfvénic flow ($M_A < 0.1$).

The steady, radial magnetic field during the interval of the Type III storm has been reconfigured by the passage of the CME. The field inversion, which occurs during the brief interval between $t_{7a}$ and $t_{7b}$, corresponds to the Carrington longitude ($300^{\circ}$) of the active region responsible for the CME (AR 13102, previously named AR 13088).

\subsection{Circularly Polarized Radio Burst Storm}

\begin{figure}[!htb]
  \plotone{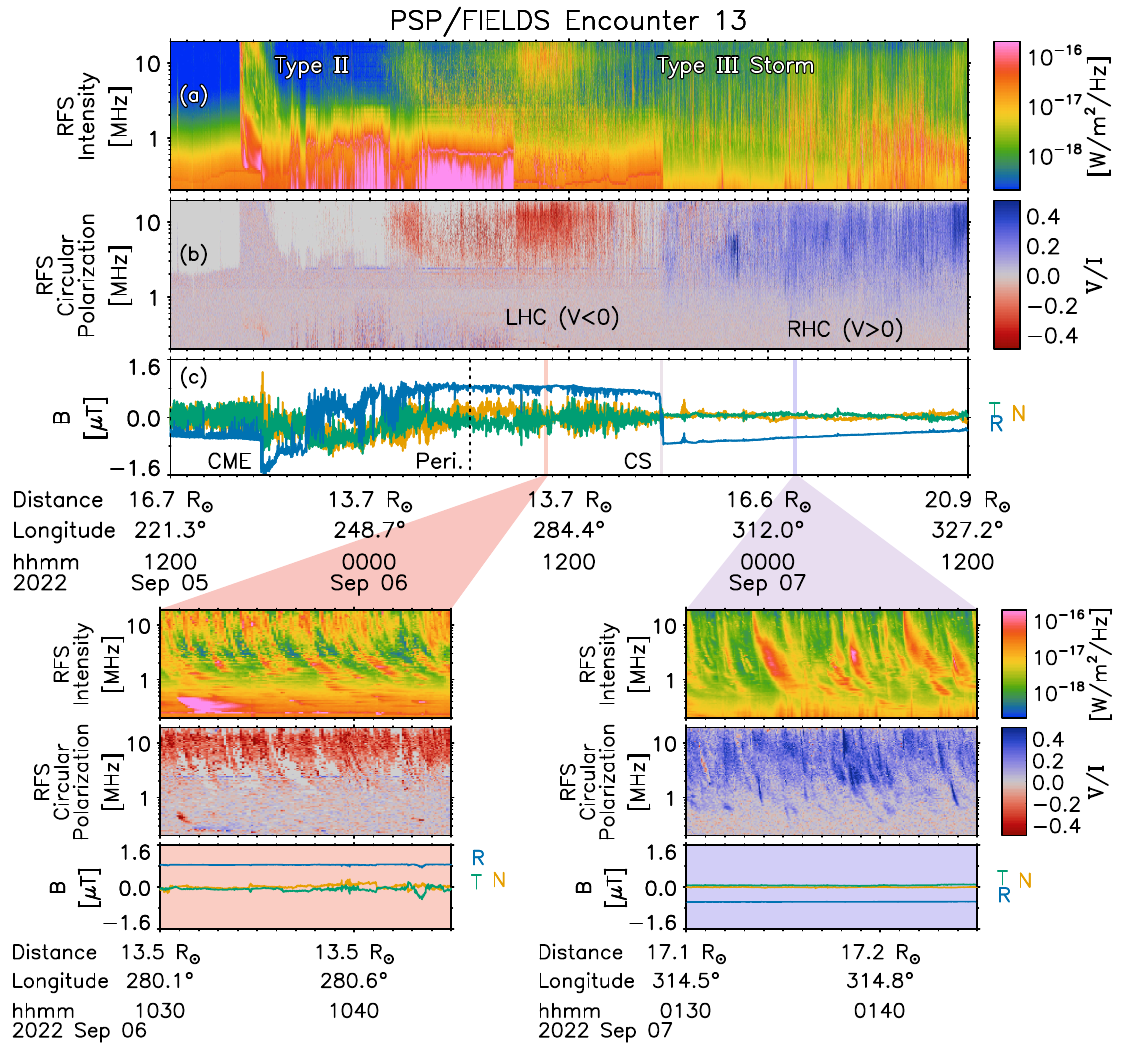}
  \caption{PSP observations of the Type III radio burst storm following the September 5 CME. The primary figure shows the intensity (a) and circular polarization (b) of the radio emission, as well as the magnetic field (c). Secondary figures show short intervals before and after crossing the current sheet, with LHC (bottom left) and RHC (bottom right) dominated emission.
    \label{fig:spectrogram}}
\end{figure}

Figure \ref{fig:spectrogram} shows \psp{} radio and magnetic field data before, during and after the September 5 CME. Figure \ref{fig:spectrogram}a shows the Stokes intensity $I$, while \ref{fig:spectrogram}b shows the Stokes circular polarization $V/I$. The intensity and polarization spectrograms include data from both the RFS HFR and LFR receivers. On the Stokes $V/I$ spectrogram, red indicates $V<0$ or LHC emission, while blue indicates $V>0$ and RHC. Circular polarization is not plotted when the intensity is negligible, for example, at high frequencies (${>}2$ MHz) during the interval before the CME.

Figure \ref{fig:spectrogram}c shows the \emph{in situ} magnetic field in RTN coordinates, illustrating the disturbed field generated by the CME and the sharp transition in radial field at the current sheet (CS) crossing. Below the primary plot of Figure \ref{fig:spectrogram}, two smaller plots show details of the storm for two representative 15 minute intervals. The left plot shows an interval before crossing the CS, and the right plot after the crossing. The steady outward and inward nature of the radial field can be seen in the $B_R$ component in each of the two smaller plots.

A few isolated Type III bursts are observed prior to the CME eruption at $t_0$, but there is no sign of continous storm emission. An intense, complex Type III radio burst and a Type II burst start at $t_0$. The bursts are signatures of electrons accelerated along open field lines (Type III) and by the CME-driven shock (Type II). No significant circular polarization is observed in the intense Type III, or in the Type II associated with the CME.

The onset of the Type III storm is gradual, without a definite starting point.  While \psp{} is the sheath region of the ICME shock after time $t_1$, sporadic Type III emission is apparent, although this period lacks the quasi-continuous behavior characteristic of storms. During this period, no dominant sense of Type III polarization is evident. After approximately 09-06/01:00, Type III emission is quasi-continuous, with a consistent LHC polarization. At the time of the CS crossing (boundaries $t_{7a}$ and $t_{7b}$, indicated in Figure \ref{fig:spectrogram}c with gray shading), the dominant sense of polarization changes from LHC to RHC, corresponding to the change of the magnetic field from outward polarity to inward polarity. The storm continues for the remainder of Encounter 13.

The circular polarization $V/I$ signal is observed at frequencies above ${\sim}$1 MHz, while the intensity $I$ of the Type III storm persists to the plasma frequency ($f_p$), which ranges from $100-300$ kHz during the storm interval shown in Figure \ref{fig:spectrogram}. The concentration of the polarized component at higher frequencies is consistent with previous observations \citep{2007SoPh..241..351R,2020ApJS..246...49P}, which demonstrate increases in polarization fraction with increasing frequency.

Polarization of fundamental plasma emission is driven by the difference between the two electromagnetic modes available for radio emission at frequencies near $f_p$, the $O$- and $X$-modes. The index of refraction for the $O$-mode allows emission to propagate from the source region to the observer, while $X$-mode emission is prevented from propagation. The propagating $O$-mode emission is inherently LHC polarized when traveling along the direction of $\mathbf{B}$, consistent with the observations in Figures \ref{fig:spectrogram} and \ref{fig:magnetic_field}.

However, IP Type III (and Type II) radio emissions are never observed to be $100\%$ polarized, as this simple picture would imply. Effects of depolarization \citep{1984SoPh...90..139W,2006ApJ...637.1113M} or of generation of emission in both $O$- and $X$-modes simultaneously in inhomogeneous plasma in the presence of density fluctuations \citep{2007PhRvL..99a5003K}, can result in incomplete or near-zero circular polarization.

The distinction between the $O$- and $X$-modes is driven by the magnetic field, and is most significant when the ratio of cyclotron frequency ($\omega_e$) to plasma frequency ($\omega_p = 2 \pi f_p$) is large. Assuming typical scaling in the solar wind, $\omega_e/\omega_p$ can reach ${\sim}0.1$ at radial distances corresponding to emission frequencies of ${\sim}1$ MHz, and rapidly declines at larger distances/lower frequencies. As the separation between modes becomes less significant, the polarization fraction also decreases. For Type III bursts occurring in regions of low density or enhanced magnetic field \citep{2024ApJ...961..136C} this effect could result in significant polarization persisting to low frequencies.

\section{Magnetic Connectivity and Circular Polarization}

\begin{figure}[!htb]
  \plotone{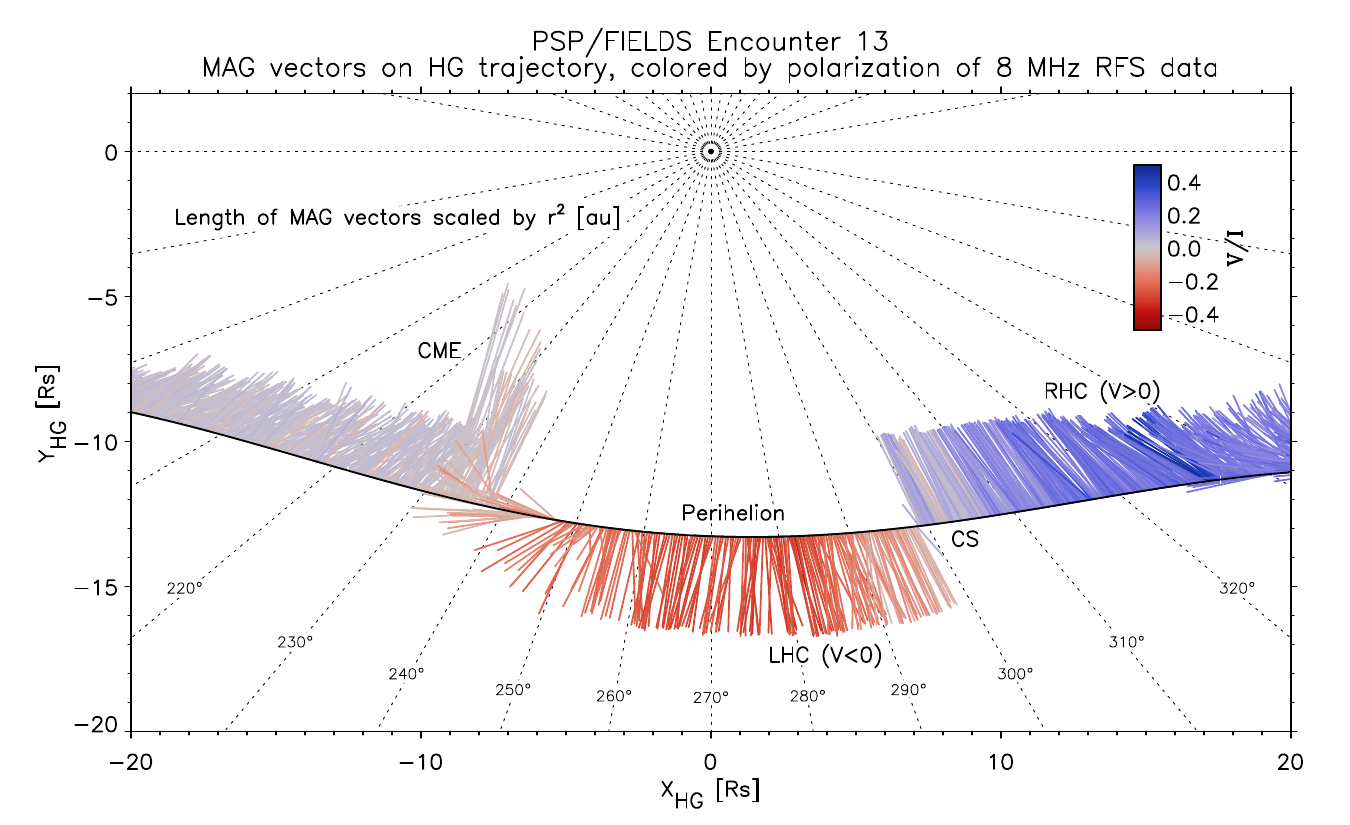}
  \caption{Trajectory of \psp{} showing magnetic field vectors and circular polarization of radio emission.
    \label{fig:magnetic_field}}
\end{figure}

Figure \ref{fig:magnetic_field} illustrates the correlation between field direction and measured polarization of radio emission as measured at \psp{}. The black line plots the trajectory of the spacecraft near perihelion of Encounter 13 in Carrington heliographic (HG) coordinates ($X_{\mathrm{HG}}-Y_{\mathrm{HG}}$). Aside from short time intervals near the CME passage and the current sheet crossing, the field is nearly radial ($|B|\sim|B_R|$).

The projection of the magnetic field vector into the $X_{\mathrm{HG}}-Y_{\mathrm{HG}}$ plane is plotted along the spacecraft trajectory. Each vector projection is scaled by $r^2$ to produce a quantity independent of radial distance, as in \citet{2021A&A...650A..18B}. With the exception of the CME and the current sheet crossing, the scaled radial field is consistent in magnitude ($B_{R}\:r^2\sim\mathrm{constant}$).

Each projected vector is also colored according to the circular polarization as measured by Stokes $V/I$ at 8 MHz. As in Figure \ref{fig:spectrogram}, this figure shows LHC emission associated with outward field, and RHC emission associated with inward field.

The magnetic configuration illustrated in Figure \ref{fig:magnetic_field} helps to explain the high degree of circular polarization observed in the Type III storm. Previous observations of polarization in RFS data have shown that polarization is an indication of direct connection to the radio source region \citep{2022ApJ...930...88N,2023A&A...674A.105D}. Figure \ref{fig:magnetic_field} is consistent with these observations, with the strongly polarized emission centered around the source region at a Carrington longitude of ${\sim}300^{\circ}$. The storm observations are also consistent with theoretical calculations that show that scattering of emission from regions of higher plasma density can destroy circular polarization \citep{2006ApJ...637.1113M}. With a dominant radial field configuration as shown in Figure \ref{fig:magnetic_field}, radio emission has a straight path from source region to observer, minimizing the possibility of reflection or scattering.

\section{Storm Statistics and Longitudinal Dependence}

Over the course of the Type III storm, well over 1000 bursts are observed. In both of the 15 minute detail plots from Figure~\ref{fig:spectrogram}, dozens of events are visible. However, it is difficult to clearly separate individual bursts, which often overlap in time and frequency, and which also vary widely in amplitude. In this section, we analyze the Type III storm in 15 minute intervals to reveal additional details about the storm-producing active region and the associated accelerated electron beams.

As discussed in the previous section, the most prominent feature of the storm is the abrupt change of emission from LHC to RHC, which indicates the change of source region from outward polarity magnetic field to inward. We therefore analyze LHC and RHC emissions separately, defining the un-normalized variables $V_{LHC}$ and $V_{RHC}$

\begin{eqnarray}
  V_{LHC} = -(V/I)&\times I&\times H(-V) \\
  V_{RHC} =(V/I)&\times I &\times H(V)
\end{eqnarray}
where $V/I$ and $I$ are the \texttt{STOKES\_V} and \texttt{PSD\_FLUX} variables from the Level 3 RFS data files, and $H$ is the Heaviside function. Separating the spectra into $V_{LHC}$ and $V_{RHC}$ allows for independent analysis of each sense of polarization, even when LHC and RHC bursts both occur in a 15 minute interval. Normalized polarization $V_{LHC}/I$ and $V_{RHC}/I$ can also be calculated for each sense. We do note that, in the case of exactly simultaneous emission, the senses of polarization may cancel each other.

\begin{figure}[!htb]
  \plotone{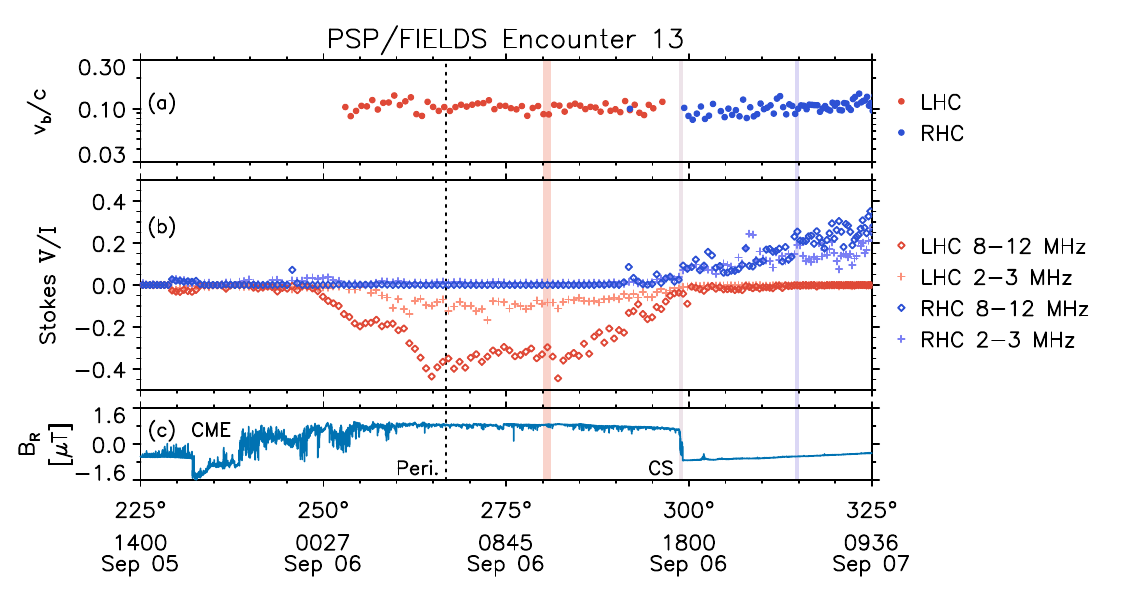}
  \caption{Derived beam speed, circular polarization, and radial magnetic field as a function of Carrington longitude for the interval shown in Figure \ref{fig:spectrogram}. \label{fig:carrington}}
\end{figure}

Figure \ref{fig:carrington} shows the beam speed and normalized polarization determined for 15 minute intervals during the CME and subsequent Type III storm. The time interval included in Figure \ref{fig:carrington} is approximately the same interval as in Figure \ref{fig:spectrogram}, but Figure \ref{fig:carrington} uses Carrington longitude as the abscissa rather than time. All LHC-specific quantities are shown in red, while RHC quantities are in blue.

Figure \ref{fig:carrington}a shows the beam speed $v_{\text{beam}}/c$, for intervals where an LHC or RHC beam speed could be determined. For each interval, $v_{\text{beam}}$ is determined using a time-lag technique developed by \citet{2012ApJ...753...35V} for use in imaging data. In the radio frequency domain, the technique has been applied to modeled radio signatures from closed loops \citep{2021ApJ...922..128C} and analysis of Type IIIs observed by \psp{} \citep{2021PhDT........15C}.

In this study, the time lag between two different frequency channels $f_i$ and $f_j$ from either the $V_{LHC}$ or $V_{RHC}$ data is calculated by identifying the time difference $t_{ij}$ which maximizes the cross-correlation between the frequency channels. The radial distance corresponding to $f_i$ and $f_j$ is calculated by first converting $f$ to electron density, assuming fundamental emission ($f = 9\times10^3 \sqrt{n_e}$). We assume the emission is fundamental because at these frequencies the fundamental component of Type III bursts tends to be more strongly polarized than the harmonic \citep{2023ApJ...955L..20J}.
While \cite{2021ApJ...922..128C} focused on emission from closed loops, we assume that the post-CME magnetic field lines are open and radial. We can then solve for $r$ from any given $f$ assuming a radial electron density profile. We use the model from \cite{1998SoPh..183..165L}, which is defined for a nominal electron density of 7.2 $\text{cm}^{-3}$ at 1 au:

\begin{equation}
  n_e = 3.3 \times 10^5 r^{-2} + 4.1 \times 10^6 r^{-4} + 8.0 \times 10^7 r^{-6}
  \label{eqn:ldb}
\end{equation}

To determine radial distances based on the \psp{} E13 measurements, we scale Equation \ref{eqn:ldb} by $(n_{\text{PSP}}/7.2{\:\text{cm}^{-3}}) \times (r_{\text{PSP}}/215{\:{R_{\sun}}})^2$, where $r_{\text{PSP}}$ is the radial distance of \psp{} and $n_{\text{PSP}}$ is the observed electron density from Level 3 RFS simplified quasi-thermal noise (SQTN) data \citep{2020ApJS..246...44M}. The radial distances $r_i$ and $r_j$ corresponding to $f_i$ and $f_j$ can then be determined by solving the scaled version of Equation \ref{eqn:ldb}. After determining the time delay and radial distance between two frequency channels, the beam speed is given by:

\begin{equation}
  v_{ij} = (r_j - r_i) / t_{ij}
  \label{eqn:beamspeed}
\end{equation}

The proximity of the spacecraft to the Sun and the radial nature of the magnetic field allow us to ignore any correction for Parker spiral effects in Equation \ref{eqn:beamspeed}.

The polarized emission occurs primarily above 1 MHz, so we calculate beam speeds using the HFR data, which contains 64 frequency channels covering a bandwidth of 1.2--19.2 MHz. After eliminating channel pairs with no significant cross-correlation, and pairs close in frequency, where the time delay would be shorter than the RFS sampling cadence, valid $v_{ij}$ pairs are averaged to produce a single overall $v_{\text{beam}}$ measurement for each 15 minute interval. When the polarized emission in a particular sense is absent or limited during a 15 minute interval, no point is plotted.

The beam speeds shown in Figure \ref{fig:carrington}a remain steady throughout the storm at $\sim$0.1$c$, a typical value for Type III emission \citep{2000GMS...119..115D}. No significant difference between the LHC (red) and RHC (blue) intervals is apparent.

Figure \ref{fig:carrington}b shows time profiles of the normalized Stokes parameter $V/I$, for two selected frequency ranges. As in Figure \ref{fig:carrington}a, LHC points are plotted in red, and RHC in blue, with each point representing a 15 minute interval of data. As discussed in the previous sections, the current sheet crossing near $300^{\circ}$ Carrington longitude separates the LHC and RHC intervals. A gradual decrease in the circular polarization fraction is apparent for both senses of polarization, for longitudes close to the crossing point. The start of the decrease in polarization fraction is most clearly apparent in Figure \ref{fig:carrington} for the LHC 8-12 MHz data, where a decrease starts between Carrington longitude between $280^{\circ}$ and $285^{\circ}$. The $20^{\circ}$ to $25^{\circ}$ difference between these longitudes and the current sheet may be indicative of the width of the emission pattern of the individual storm bursts.

Figure \ref{fig:carrington}c shows the radial component of the magnetic field, as a reference point for comparison between Figure \ref{fig:carrington} and the preceding figures. The red, gray, and blue highlighted intervals of Carrington longitude in Figure \ref{fig:carrington} correspond to the highlighted time intervals in Figure \ref{fig:spectrogram}.

\section{Discussion}

From the radio and \emph{in situ} observations, a picture of the CME and Type III storm emerges. The passage of the CME disturbs the magnetic field in the corona. The post-CME field is highly radial, with a sharp boundary between inward and outward open field lines. On either side of the active region boundary, the configuration of the magnetic field is favorable for quasi-continous electron acceleration, and the resulting electron beams produce the Type III storm. The Type III emission is directed out along radial field lines, with minimal reflection and scattering. The emission is polarized in the sense consistent with $O$-mode emission at the fundamental of the local plasma frequency in the source region. The low level of scattering preserves the separation between LHC and RHC intervals, and also preserves the polarization signal itself.

Previous coordinated solar flare observations have emphasized the correlation between Type III radio bursts and other flare-related phenomena such as X-ray emission and coronal jets \citep{2017LRSP...14....2B,2016SSRv..201....1R}. Individual, non-storm Type III bursts can be directly associated with EUV jet observations and X-ray emission \citep[e.g.][]{2009A&A...508.1443B,2011A&A...531L..13I,2011ApJ...742...82K,2016A&A...589A..79M,2018NatSR...8.1676C}.

For large events, reconnection can reorganize the topology of the magnetic field, releasing a large fraction of the stored pre-flare energy in the form of radiation and energetic particles. Smaller-scale energy release can enable recurrent flare and radio burst activity, with each individual release small enough that the overall field configuration is not significantly disturbed. Examples of this type of recurrent activitiy, with Type III bursts accompanied by EUV observations, have been observed by \cite{2019ApJ...873..110P} and \cite{2021A&A...650A...6C}.

These observations demonstrate that the source of each phenomena is energy release driven by magnetic reconnection. Type III storms are also believed to be produced by this same fundamental process. The direct relation between storm radio emission and other phenomena are less clear, likely due to the smaller nature of the individual bursts in the storm. \citet{2007ApJ...657..567M} demonstrated that the distribution of intensity of storm bursts (``micro-type III'' in the terminology of \citet{2007ApJ...657..567M}) is distinct from those of individual bursts, and found no relation between storms and soft X-ray emission.

\cite{2007ApJ...657..567M} suggest that Type III storms are generated near the boundary between an active region and the open field lines of a neighboring coronal hole. \cite{2011A&A...526A.137D} proposed a model where the continuous burst activity of a Type III storm is driven by active region expansion, which in turn drives interchange reconnection and acceleration of electron beams along these open field lines. \cite{2011A&A...526A.137D} demonstrated the consistency of this model with EUV observations of coronal outflows.

While Type III storms are direct signatures of electron beams, the persistent reconnection that drives the beams and radio emission can also accelerate ions. \cite{2015SoPh..290.2423T} proposed that type III storm activity may be a precursor of a CME with high levels of solar energetic particles (SEPs), with the seed particles required for a high SEP event \citep{2013ApJ...770...73L} generated by the same magnetic field configuration that produces the storm.

\cite{9814233} studied two similar CMEs, one which was preceded by a Type III storm and one which was not. The storm-associated CME indeed showed a significantly higher flux of SEPs, supporting the suggestion that storms are associated with seed particle generation. \cite{9814336} also demonstrated that storms are associated with higher fluxes for corotating interaction region (CIR) events, suggesting that seed particles generated at the source of the storm may travel to the CIR region along open field lines, and there be accelerated to MeV energies. Since storm emission can be observed for many hours or days prior to a CME release, the association between storms and energetic particles could potentially result in improved predictions for SEP levels from a given eruptive active region.

\section{Conclusions}

Using a Type III radio burst storm observed by PSP, we demonstrate a clear association of the sense of circular polarization with the \emph{in situ} magnetic field.  The sense and degree of circular polarization is consistent with fundamental emission in the $O$-mode. The beam speed, derived from the frequency drift of the Type III storm emission, is steady throughout the duration of the event, consistent with a stable magnetic configuration generating quasi-continous acceleration of electrons.

For these particular observations, the proximity of the spacecraft to the Sun and the quiet radial field observed during the storm allow for a straightforward connection to the inferred source region. The close agreement between the \emph{in situ} and radio observations shows that Type III storm observations can be used to remotely probe the magnetic configuration of the source active region.

Recent studies have indicated that Type III storms may be associated with generation of seed particles, which can increase SEP fluxes from CIRs and CMEs. This correlation suggests that Type III storm observations could help to predict the likelihood that a solar transient will be associated with high fluxes of energetic particles.

\begin{acknowledgements}
Parker Solar Probe was designed, built, and is now operated by the Johns Hopkins Applied Physics Laboratory as part of NASA’s Living with a Star (LWS) program under NASA contract NNN06AA01C. Data access and processing was done using SPEDAS \citep{2019SSRv..215....9A}. FIELDS data products are publicly available at the FIELDS SOC webpage, https://fields.ssl.berkeley.edu.
\end{acknowledgements}

\bibliography{rfs_storm_papers,rfs_storm_papers_no_ads}{}
\bibliographystyle{aasjournal}


\end{document}